\documentclass[aip,jmp,twocolumn,superscriptaddress,amsmath,amssymb]{revtex4}

\usepackage{graphicx}
\usepackage{latexsym}    %
\usepackage{color}       %
\usepackage{dcolumn}     
\usepackage{bm}          
\usepackage{amssymb}     
\usepackage{hyperref}    
\expandafter\let\csname equation*\endcsname\relax
\expandafter\let\csname endequation*\endcsname\relax
\usepackage{amsmath}
\usepackage{mathtools}          

\newcommand{\be}{\begin{equation}}
\newcommand{\ee}{\end{equation}}
\def\LB#1{{\textcolor{black}{#1}}}    

\newcommand{\bn}{\boldsymbol{n}}
\newcommand{\bq}{\boldsymbol{q}}
\newcommand{\bw}{\boldsymbol{w}}
\newcommand{\bx}{\boldsymbol{x}}
\newcommand{\by}{\boldsymbol{y}}
\newcommand{\hv}{\hat{v}}

\newcommand{\ve}[1]{\ensuremath{\boldsymbol{#1}}}
\newcommand{\ma}[1]{\ensuremath{\mathbb{#1}}}

\begin{document}

\title{Zermelo's problem: Optimal point-to-point navigation in 2D turbulent flows using Reinforcement Learning \footnote{article submitted to AIP Publishing Chaos, Focus Issue: When Machine Learning Meets Complex Systems: Networks, Chaos and Nonlinear Dynamics (2019)}}

\author{L. Biferale}
\affiliation{Dept. Physics and INFN University of Rome Tor
vergata, via della Ricerca Scientifica 1, 00133 Rome, Italy.}
\author{F. Bonaccorso}
\affiliation{Dept. Physics and INFN University of Rome Tor
vergata, via della Ricerca Scientifica 1, 00133 Rome, Italy.}
\affiliation{Center for Life Nano Science@La Sapienza, Istituto Italiano di Tecnologia, 00161 Roma, Italy}
\author{M. Buzzicotti}
\affiliation{Dept. Physics and INFN University of Rome Tor
vergata, via della Ricerca Scientifica 1, 00133 Rome, Italy.}
\author{P. Clark Di Leoni}
\affiliation{Dept. Physics and INFN University of Rome Tor
vergata, via della Ricerca Scientifica 1, 00133 Rome, Italy.}
\affiliation{Dept. of Mechanical Engineering, Johns Hopkins University, Baltimore, Maryland 21218, USA.}
\author{K. Gustavsson}
\affiliation{Dept. of Physics, University of Gothenburg, Gothenburg, 41296, Sweden.}

\date{\today}

\begin{abstract}
To find the path that minimizes the time to
navigate between two given points in a fluid flow is known as Zermelo's
problem. Here, we investigate it by using a
 Reinforcement Learning (RL) approach for the case of a vessel which has a slip velocity with fixed intensity, $V_{\rm s}$,
but variable direction and navigating in a 2D turbulent sea.  \LB{We show that  an Actor-Critic RL algorithm is able to find quasi-optimal solutions for both time-independent and chaotically evolving flow configurations. For the frozen case,}  we also compared the results
with strategies obtained analytically from continuous Optimal Navigation
(ON) protocols.  We show that for our application, ON solutions are unstable for the typical duration of the navigation process, and are therefore not useful in
practice. On the other hand, RL solutions are much more robust with
respect to small changes in the initial conditions and to external
noise, even when $V_{\rm s}$ is much
smaller than the maximum flow velocity.  Furthermore, we show
how the RL approach is able to take advantage of the flow properties in order to reach the target, especially when the steering speed is small.
\end{abstract}
\maketitle

{\bf Zermelo's point-to-point optimal navigation problem in the presence of a  complex flow is key for a variety of geophysical and applied instances.  In this work, we apply  Reinforcement Learning to solve Zermelo's problem in a multi-scale 2d turbulent snapshot \LB{for both frozen-in-time velocity configurations and fully time-dependent flows}. We show that our approach is able to find the quasi-optimal path to navigate from two distant points with high efficiency, even comparing with policies obtained from  optimal control theory.  Furthermore, we connect the learned policy with the topological flow structures that must be harnessed by the vessel to navigate fast. Our result can be seen as a first step  towards more complicated applications to surface oceanographic problems and/or 3d chaotic and turbulent flows. }

\section{Introduction}
\begin{figure*}
\includegraphics[scale=0.6]{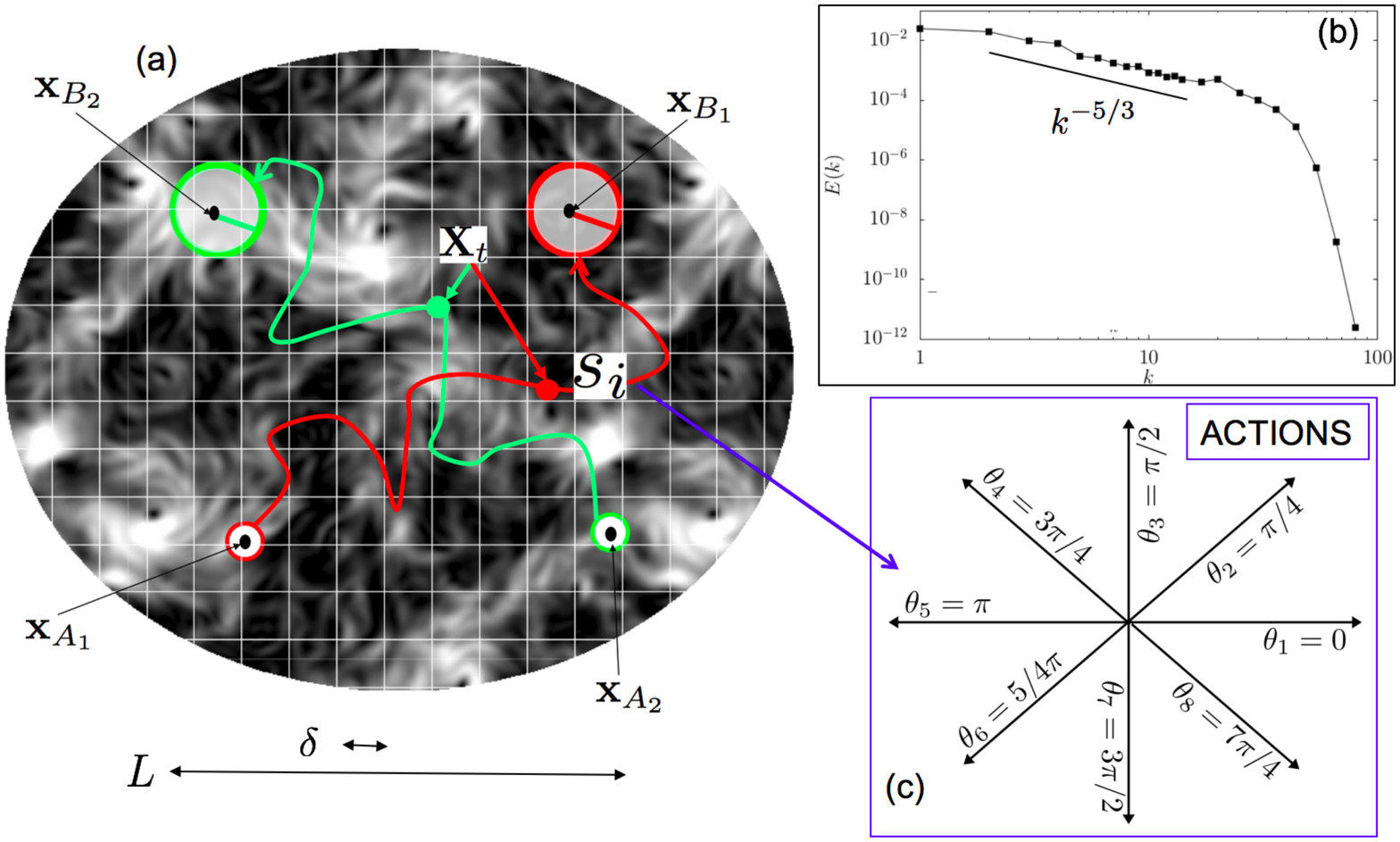}
\caption{Panel a: Image of the turbulent snapshot used as the advecting
flow, with the starting and ending points of the two considered problems.
Problem P1 goes from ${\ve x}_{A_1} \to \bx_{B_1}$ (red) and problem
P2 goes from ${\ve x}_{A_2} \to \bx_{B_2}$ (green). For each case, we also show an
illustrative trajectory $\ve X_t$.  The flow is obtained from a
fully periodic snapshot of a 2D turbulent configuration in the inverse energy
cascade regime with a multi-scale power-law  Fourier spectrum, $E(k) =
\sum_{k< {\ve k} <k+1} |{\ve u}({\ve k})|^2 \sim k^{-5/3}$, shown in
panel (b). For RL optimization, the initial conditions are taken randomly
inside a circle
of radius $d_{A_i}$ centered around ${\ve x}_{A_i}$. Similarly, the final targets are the
circles of radius $d_{B_i}$ centered around ${\ve x}_{B_i}$ for each
problem.
The flow area is covered by tiles $s_i$ with $i=1,\dots,N_s$ and
$N_s=900$ of size $\delta\times\delta$ which identify the state-space
for the RL protocol. Every time interval $\Delta t$, the agent selects one of the 8
possible actions $a_j$ with $j=1\dots 8$ (the steering directions $\theta_j$ depicted in panel
(c)) according to the policy $\pi(a|s)$ (see
Sec.~\ref{sec:RL_methods}), where $\pi$ is the probability distribution of the action $a$ given the current state $s$ of the agent at that time. The large-scale periodicity of the underlying flow is $L$, and we fixed $\delta=L/10$.
}
\label{fig:initial}
\end{figure*}

Path planning for small autonomous marine vehicles
\cite{petres2007path,witt2008go} such as wave and current gliders
\cite{kraus2012wave,smith2011predicting}, active drifters
\cite{lumpkin2007measuring,niiler20011}, buoyant underwater explorers, and small swimming drones is key for many
geo-physical \cite{lermusiaux2017future} and engineering
\cite{bechinger2016active, kurzthaler2018probing, popescu2011pulling,
baraban2012transport} applications. In nature, these vessels are
affected by environmental disturbances like wind, waves and ocean
currents, often in competition, and often characterized by unpredictable
(chaotic) evolutions. This is problematic when one wants to send probes
to specific locations, for example when trying to optimize data-assimilation for environmental applications \cite{lermusiaux2017future,carrassi2018data,lakshmivarahan2013nudging,
leoni2018unraveling, leoni2019synchronization}.  Most of the times, a
dense set of fixed platforms or manned vessels are not economically
viable solutions. As a result, scientists rely on networks of moving
sensors, e.g. near-surface currents drifters
\cite{centurioni2018drifter} or buoyant explorers
\cite{roemmich2009argo}. In both cases, the platforms move with
the surface flow (or with a depth current) and are either fully passive
\cite{centurioni2018drifter} or inflatable/deflatable with some
predetermined scheduled protocol \cite{roemmich2009argo}. The main
drawback is that they might be distributed in a non-optimal way, as they
might accumulate in uninteresting regions, or disperse away from key
points. Beside this {\it applied} motivation, the problem of (time)
optimal point-to-point navigation in a flow, known as Zermelo's problem
\cite{zermelo1931}, is interesting {\it per se} in the framework of Optimal
Control Theory \cite{bryson2018, ben2010optimal, liebchen2019optimal,
hays2014route}.

In this paper, we tackle Zermelo's problem for a specific but important
application, the case of a two-dimensional fully
turbulent flow\cite{xia2011upscale,boffetta2012two,alexakis2018cascades} with an
entangled distribution of complex spatial features, such as recirculating eddies or shear-regions,  and with multi-scale
spectral properties (see Fig.~\ref{fig:initial} for a graphical summary
of the problem).  In such conditions, \LB{even for time-indenpendet flow configurations}, trivial or naive navigation policies
 can be extremely inefficient and ineffective.
To overcome this, we have implemented one approach based on Reinforcement
Learning (RL) \cite{sutton2018} in order to find a robust quasi-optimal policy that accomplish the task, \LB{and we applied it to time-independent flow configurations and to the case when the underlying flow evolves with its own chaotic and turbulent dynamics given by the Navier-Stokes equations.}   Furthermore, we compare RL with an approach based on  Optimal Navigation
(ON) theory \cite{rugh1996linear, pontryagin2018mathematical}. To the
best of our knowledge, only simple advecting flows have been studied so
far for both  ON \cite{techy2011optimal, yoo2016path,
liebchen2019optimal} and RL \cite{yoo2016path}.

Promising results have been obtained when applying RL algorithms to similar problems, such as the training of smart inertial particles or swimming particles navigating intense vortex regions~\cite{colabrese2018smart}, Taylor Green flows~\cite{colabrese2017flow} and ABC flows \cite{gustavsson2017finding}. RL has also been successfully implemented to reproduce schooling of fishes \cite{gazzola2016learning,verma2018efficient}, soaring of birds in a turbulent environments \cite{reddy2016learning,reddy2018glider} and in many other applications \cite{muinos2018reinforcement,novati2018deep,tsang2018self}. Similarly, in the recent years, artificial intelligence techniques are establishing themselves
as new data driven models for fluid mechanics in general \cite{pathak2017using,king2018deep,vlachas2018data,lu2018attractor,mohan2019compressed,brunton2019machine}.

In this paper, we show that for the case of vessels that
have a slip velocity with fixed intensity but variable direction, RL can
find a set of quasi-optimal paths to efficiently navigate the flow.  Moreover,
RL, unlike ON, can provide a set of highly stable solutions, which are
insensitive to small disturbances in the initial condition and
successful even when the slip velocity is much smaller than the guiding
flow. We also show how the RL protocol is able to take advantage of
different features of the underlying flow in order to achieve its task,
indicating that the information it learns is non-trivial.

The paper is organized as follows. In Sec.~\ref{sec:setup} we present
the general set-up of the problem, write the equations of motion of the
vessels used, we give details on the underlying flow and tasks. In
Sec.~\ref{sec:RL} we first present an overview of the RL algorithm used
in this paper and then show the results obtained using it, while in
Sec.~\ref{sec:ON} we do the same for the ON case. In
Sec.~\ref{sec:comparison} we compare the results obtained from the two
approaches. Finally, we give our conclusions in
Sec.~\ref{sec:conclusions}.

\section{Problem set-up}
\label{sec:setup}
For our analysis we use one static snapshot from a numerical realization of 2D turbulence, and
try to learn the optimal path connecting two different sets of starting
and ending points; we call these problems P1 and P2, respectively. In
Fig.~\ref{fig:initial} we show a sketch of the set-up (see the caption of the figure  for further details
on the turbulent realization used and how it was generated). Our goal is to find (if they
exist) trajectories that join the region close to $\bx_{A_1}$ with a
target close to $\bx_{B_1}$ (problem P1) and $\bx_{A_2}$ with $\bx_{B_2}$
(problem P2) in the shortest possible time assuming that the vessels obey
the following equations of motion:
\begin{equation}
\begin{cases}
  \dot {\ve X}_t  = \LB{{\ve u}({\ve X}_t,t)} + {\ve U}^{ctrl}({\ve X}_t) \\
  {\ve U}^{ctrl}({\ve X}_t) = V_{\rm s} {\bn}({\ve X}_t)
\label{eq:theta}
\end{cases}
\end{equation}
where \LB{${\ve u}({\ve X}_t,t)$} is the velocity of the underlying 2D
advecting flow, and $${\ve U}^{ctrl}({\ve X}_t) = V_{\rm s}{\bn}({\ve X}_t)$$ is the control slip
velocity of the vessel with fixed intensity $V_{\rm s}$ and varying
steering direction ${\bn}$:
$$ {\bn} ({\ve X}_t) = ( \cos[\theta_t], \sin[\theta_t] )\,,$$
where the angle is evaluated along the trajectory, $\theta_t = \theta({\ve X}_t)$.
We introduce a dimensionless slip velocity by dividing $V_{\rm s}$ with the maximum velocity $u_{\max}$ of the underlying flow:
\begin{equation}
\tilde{V}_{\rm s} = V_{\rm s}/u_{\max}\,.
\end{equation}
In this framework, Zermelo's problems reduces to optimize the time-space
dependency of $\theta$ in order to reach the target~\cite{zermelo1931}. \LB{For the case when the flow is time independent or its evolution is so slow that it can be considered approximately frozen}, a general  solution, given by optimal control theory, can be found in Refs.~\cite{techy2011optimal,mannarini2016visir}. In particular, assuming that
the angle $\theta$ is controlled continuously in time, one can prove
that, if there exists an optimal trajectory that joins $\bx_A$ with
$\bx_B$ with a given  initial angle $\theta_{t_0}$, the optimal steering
angle must satisfy the following time-evolution:
\begin{equation}
\dot \theta_t = A_{21}\sin^2\theta_t - A_{12}\cos^2\theta_t + (A_{11} -
A_{22})\cos\theta_t \sin\theta_t\,, \label{eq:ON}
\end{equation}
where $A_{ij}=\partial_ju_i(\ve X_t)$ is evaluated along the agent
trajectory $\ve X_t$ obtained from Eq.~(\ref{eq:theta}). The set of equations (\ref{eq:theta}) together with
(\ref{eq:ON}) form a three-dimensional dynamical system, which may result in chaotic dynamics even though the fluid velocity is 2D and time-independent (tracer particles cannot exhibit chaotic dynamics in such flow). Due to the sensitivity to small errors in chaotic systems the ON approach might become useless for many practical applications.
Moreover, even in the presence of a global non-positive maximal Lyapunov
exponent, where the long time evolution of a generic trajectory is
attracted toward fixed points or periodic orbits for almost all initial
conditions, the finite time Lyapunov exponents (FTLE)
\cite{ott2002chaos,angelo2009chaos} can be positive for particular
initial conditions and for a time longer than the typical navigation time.
In this case, a navigation protocol based on (\ref{eq:ON}) would
be unstable for all practical purposes. This is most likely the reason why 
previous works on ON have dealt mainly with simple advecting flow configurations \cite{techy2011optimal, yoo2016path, liebchen2019optimal}.

\section{The Reinforcement Learning approach}
\label{sec:RL}

\subsection{Methods}
\label{sec:RL_methods}
RL applications \cite{sutton2018} are based on the idea that an optimal solution for certain complex problems can be obtained by learning from continuous interactions of an agent with its environment. The agent interacts with the environment by sampling its states $s$, performing actions $a$ and collecting rewards $r$.
In the approach used here, actions are chosen randomly with a probability that is given by the policy function,
$\pi(a|s)$, given the current state $s$ of the surrounding environment. The goal is to find the optimal policy $\pi^*(a|s)$ that maximizes the total reward,
\begin{equation}
r_{\rm tot} = \sum_t r_t\,,
\label{eq:tot_reward}
\end{equation}
accumulated along one episode, i.e. one trial. 
To accomplish this goal, RL works in an iterative fashion. Different attempts, or episodes, are performed and the policy
is updated to improve the total reward.  The initial policy of each episode coincides
with the final policy of the previous episode. During the training phase optimality is approached as the total reward for each episode converges (as a function of the number of episodes) to a fixed value (up to stochastic fluctuations).

In our case the vessel acts as the agent and the two-dimensional flow as the environment.
As shown in Fig.~\ref{fig:initial}, we define the states by covering the flow domain with square tiles
$s_i$ with $i=1,2,\dots,N_s$ of size $\delta\times\delta$. Here $N_s=900$ and $\delta \sim L/10$, where {$L$ is the
large-scale periodicity of the flow.}
In other words, we suppose that the agent is able to identify its absolute
position in the flow within a given approximation determined by the tile size $\delta$.
Furthermore, to be realistic,  we
allow the agent to sample states and change action only at given time intervals $\Delta t \sim T_v/20$, where $T_v$ is the characteristic  flow
time, $T_v=L/u_{\max}$. The possible actions (steering directions) correspond to the eight angles shown in
Fig.~\ref{fig:initial}, namely $\theta_j = (j-1)\pi/4$ with $j=1,\dots,8$.
Each episode is defined as one attempt to reach the target, where we make sure that the sum
in Eq.~\eqref{eq:tot_reward} is always finite by imposing a maximum time
$T_{\max}$ (chosen to be of the order of $10$ times the typical navigation time) after which we terminate the episode.
To identify a time-optimal trajectory we {use a potential based reward shaping \cite{art1} at each time $t$ during the learning process:} 
\begin{equation}
    r_t = -\Delta t + \frac{| {\ve x}_B- {\ve X}_{t-\Delta t}|}{V_{\rm s}} -
    \frac{| {\ve x}_B - {\ve X}_t|}{V_{\rm s}}\,,
    \label{eq:reward}
\end{equation}
where ${\ve x}_B$ is the center of the final target region.  The first term
in the RHS of Eq.~(\ref{eq:reward}) {is a contribution that accumulate to a large penalty if it takes long for the agent to reach the end point. The second and third terms give the relative improvement in the distance-from-target potential  during the
training episode, which is known to preserve the optimal policy and help the algorithm to converge faster \cite{art1}.}  An episode is finalized when the trajectory reaches the
circle of radius $d_B\sim 0.9 \, \delta$ around the target, $d_B$ is roughly $10$ times smaller than the distance between the target and the starting position, see Fig.~\ref{fig:initial}.
If an agent does not reach the target within time $T_{\max}$, or gets as
far as $3L$ from the target the episode is ended and a new episode begins. In the latter case the agent receives an extra negative reward equal to $-2 T_{\max}$, in order to strongly penalize these failures.
By summation of Eq.~(\ref{eq:reward}) {over the entire duration of the episode,} the total reward~(\ref{eq:tot_reward}) becomes
\begin{equation}
r_{\rm tot} = -T_{A\to B} + \frac{| {\ve x}_B- {\ve X}_{t_0}|}{V_{\rm s}}  - \frac{d_B}{V_{\rm s}}\,.
\label{eq:rtot_evaluated}
\end{equation}
Eq.~(\ref{eq:rtot_evaluated}) is approximately equal to the difference between the time to reach the target without a flow and the actual time taken by the trajectory: $r_{\rm tot} \approx T^{\rm free}_{A\to B}-T_{A\to B}$, where the free-flight time is defined as
\begin{equation}
T^{\rm free}_{A\to B}=\frac{| {\ve x}_B- {\ve x}_A|}{V_{\rm s}}\,.
\label{eq:Tfree}
\end{equation}
In order to converge to policies that are robust against small perturbations of the initial
condition, which is an important property in the presence of chaos, each episode is started with a uniformly random position within a given radius from the starting point $|  {\ve x}_A- {\ve X}_{t_0} | < d_A \sim 0.4 \, \delta$. Following from our action-state space discretization, with $i=1,\dots, N_s$ states  and $j=1,\dots, N_a$
actions, a natural choice for the policy parametrization is the {\it
softmax} distribution defined as:
\begin{equation}
    \pi(a_j|s_i;\bq)= \frac{\exp{h(s_i,a_j,\bq)}}{\sum_{k=1}^{N_a}
    \exp{h(s_i,a_k,\bq)}}\,,
    \label{eq:softmax}
\end{equation}
where $h(s_i,a_j,\bq) = \sum_{i'j'}q_{j',i'} \,\,
z_{i',j'}(s_i,a_j)$ is a linear combination of a $N_a\times N_s$ matrix, $\bq$,
of free parameters.
{Here we adopt the simplest choice of the feature matrix $z_{i',j'} (s_i,a_j)$: a perfect non overlapping tiling of the action-state space, $z_{i',j'} =  \delta_{i,i'}\delta_{j,j'}$.} 
Unless the matrix of coefficients converges
to a singular distribution for each state $s_i$, the softmax expression 
(\ref{eq:softmax})  leads to a stochastic dynamics even for the optimal policy.

During the training phase of the RL protocol, one needs to
estimate the expected total future reward (\ref{eq:tot_reward}). In this paper, we follow  the one-step actor-critic
method \cite{sutton2018} based on a gradient ascent in the policy parametrization. The critic approach circumvents the need to generate
a big number of trial episodes by introducing the estimation of the
the state-value function, $\hv(s_i,\bw)$;
\begin{equation}
    \hv(s_i,\bw) = \sum_{i'=1}^{N_s} w_{i'} \,\, y_{i'}(s_i),
    \label{eq:liear_valuefunct}
\end{equation}
where $y_{i'}(s_i) = \delta_{i',i}$ and $w_i$ are a set of free
parameters (similar to $q_{ji}$). The expression $\hv(s_i,\bw)$ is used to estimate the future expected reward, $\hat{r}_t$, in the gradient ascent algorithm:
\begin{equation}
\hat{r}_{t+\Delta t} = r_{t+\Delta t} + \hv(s_{t+\Delta t}, \bw)\,.
\end{equation}

Finally, the parameterizations of the policy and the state-value functions are updated every time the state-space is sampled as:
\be
\begin{cases}
    \bq_{t+\Delta t} = \bq_t + \alpha_{t} \beta_t \nabla_{\bq}
        \ln(\pi(a_t|s_t,\bq_t))
    \\
    \bw_{t+\Delta t} = \bw_t + \alpha_{t}' \beta_t \nabla_{\bw} \hv(s_t,\bw_t)
\end{cases}
\hspace{-0.3cm},
\label{eq:one-step-ac}
\ee
where $s_t,a_t$ are the $(i,j)$ state-action pairs that are explored at
time $t$ during the episode, while $\beta_t$ is the future expected
reward minus the state-value function, now used as baseline
$$
    \beta_t = [\hat{r}_{t+\Delta t} - \hv(s_t,\bw_t)]\,.
$$

The main appeal of the one-step actor-critic algorithm is that replacing
the total reward with the one-step return plus the learned state-value
function, leads to  a fully local -in time- evolution of the gradient
ascent.  The learning rates $\alpha_t$, $\alpha_t'$, follow the Adam
algorithm \cite{kingma2014adam} to improve the convergence performance over standard
stochastic gradient descent. Both gradients in
Eqs.~\eqref{eq:one-step-ac} can be computed explicitly and the one-step
actor-critic algorithm becomes
\be
\begin{cases}
\bq_{t+\Delta t} = \bq_t + \alpha_t \beta_t [{\bf z} (s_t,a_t) -
\sum_{j=1}^{N_a} \pi(a_j|s_t,\bq_t) {\bf z}(s_t,a_j)] \\
\bw_{t+\Delta t} = \bw_t + \alpha_{t}' \beta_t  \by(s_t) \nonumber
\end{cases}
\hspace{-0.3cm}.
\ee


\begin{figure}
    \includegraphics[scale=0.29]{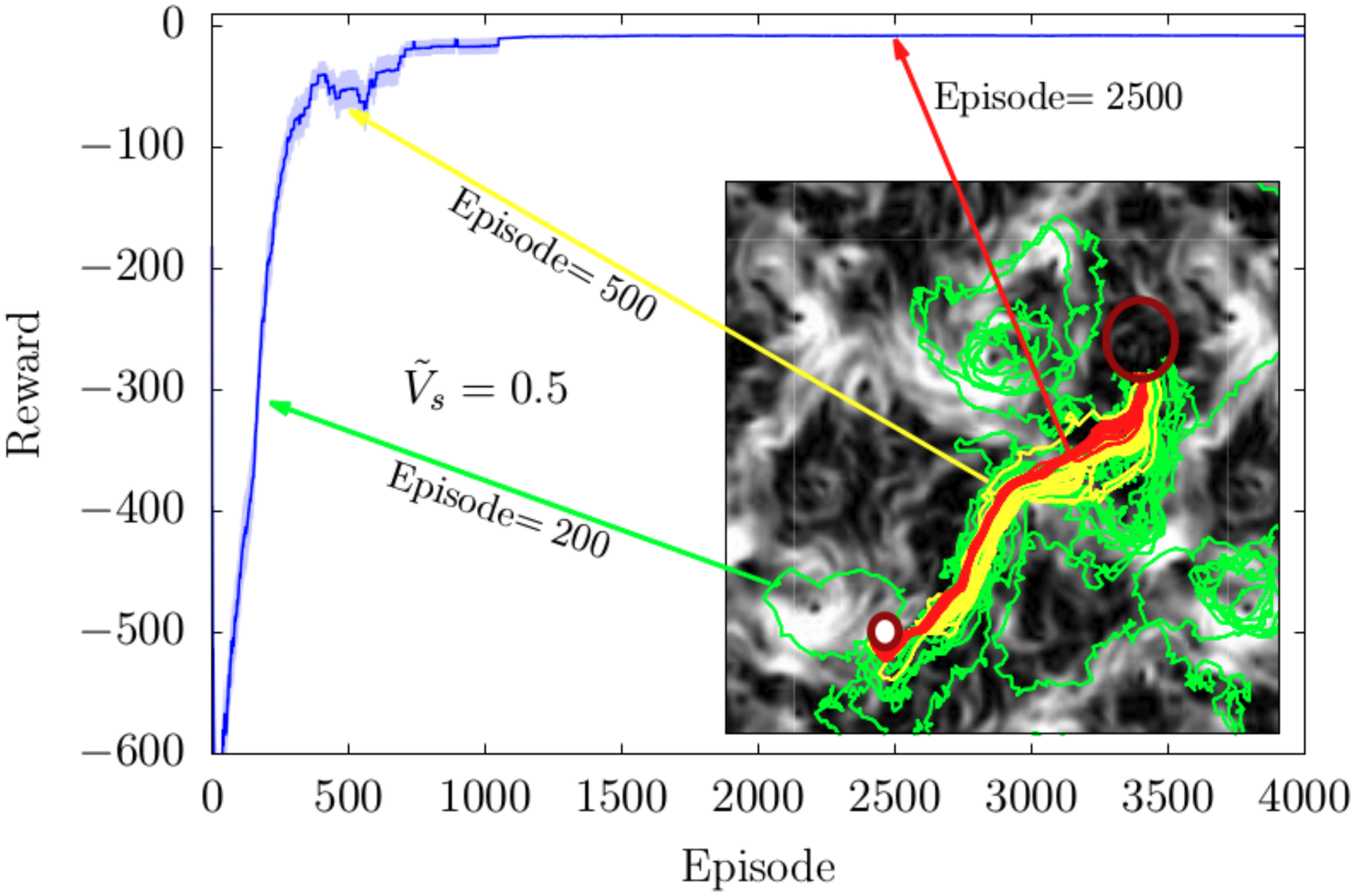}
    \caption{Evolution of the total reward $r_{\rm tot}$ averaged over a window of $150$ consecutive  episodes versus the number of
    training episodes for $\tilde V_{\rm s} = 0.5$. {The shaded area around the main curve indicates the standard deviation inside each averaged window}. Inset: evaluation of the
    policies obtained during the training at three different stages, after
    200 episodes obtaining a total reward $r_{\rm tot}\sim -300$ (green), after 500 episode $r_{\rm tot}\sim -50$ (yellow) and for a policy
    already converged to the maximum reward after 2500 episodes with the total reward increased up to $r_{\rm tot}\sim -8$ (red).
    \label{fig:training}}
\end{figure}
\begin{figure}%
    \includegraphics[scale=0.3]{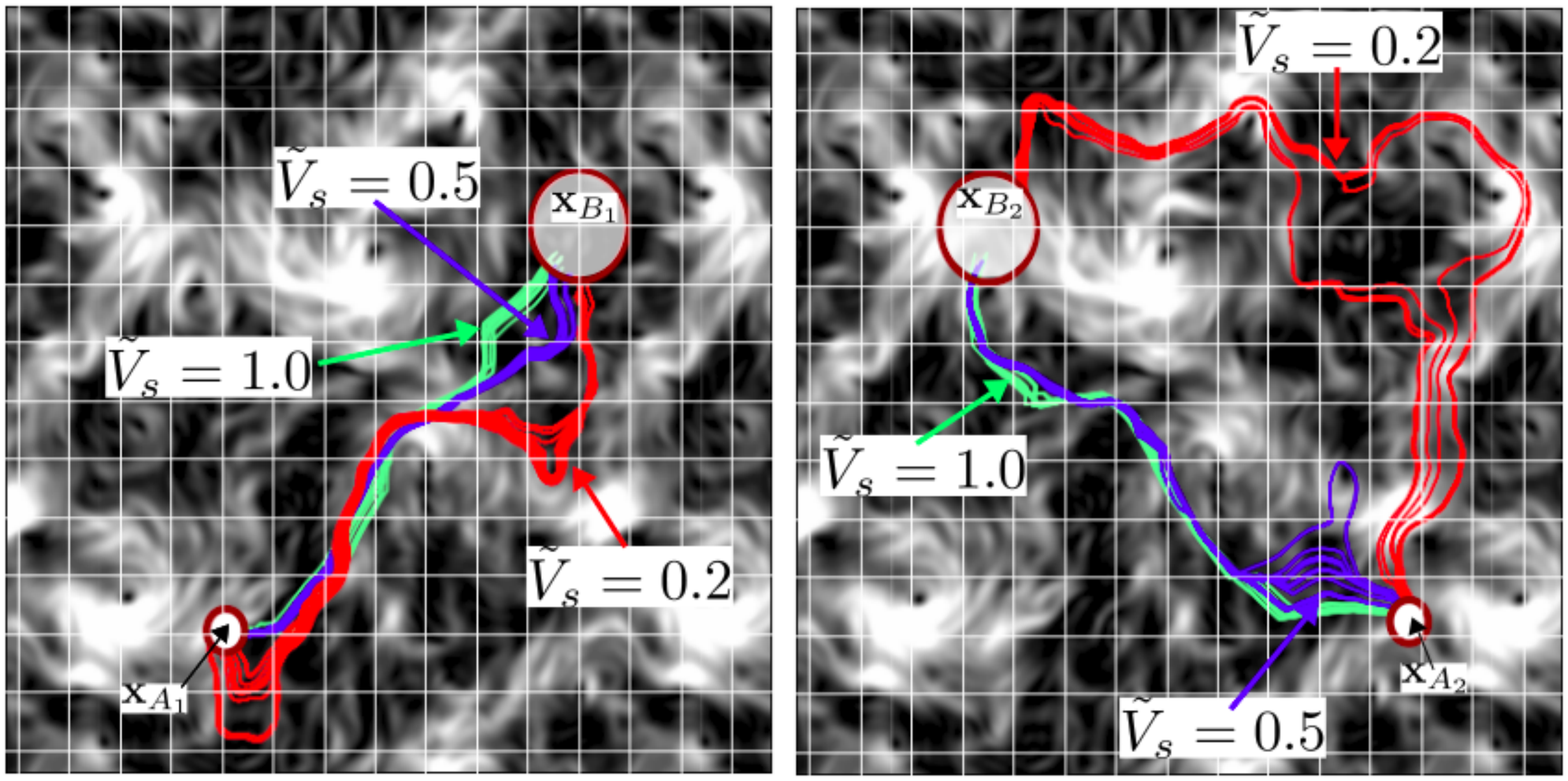}%
    \caption{\LB{Time-independent case:} examples of different trajectories generated using the
    final policy $\pi^*(a|s; \bq)$ resulting from the RL protocol, for
    different $\tilde V_{\rm s}$ and for problems P1 (left) and P2 (right). For
    each case we plot ten trajectories starting randomly in the circle around $\bx_{A_1}$  and $\bx_{A_2}$. \label{fig:comp}
 }
\end{figure}
\begin{figure}%
    \includegraphics[scale=0.32]{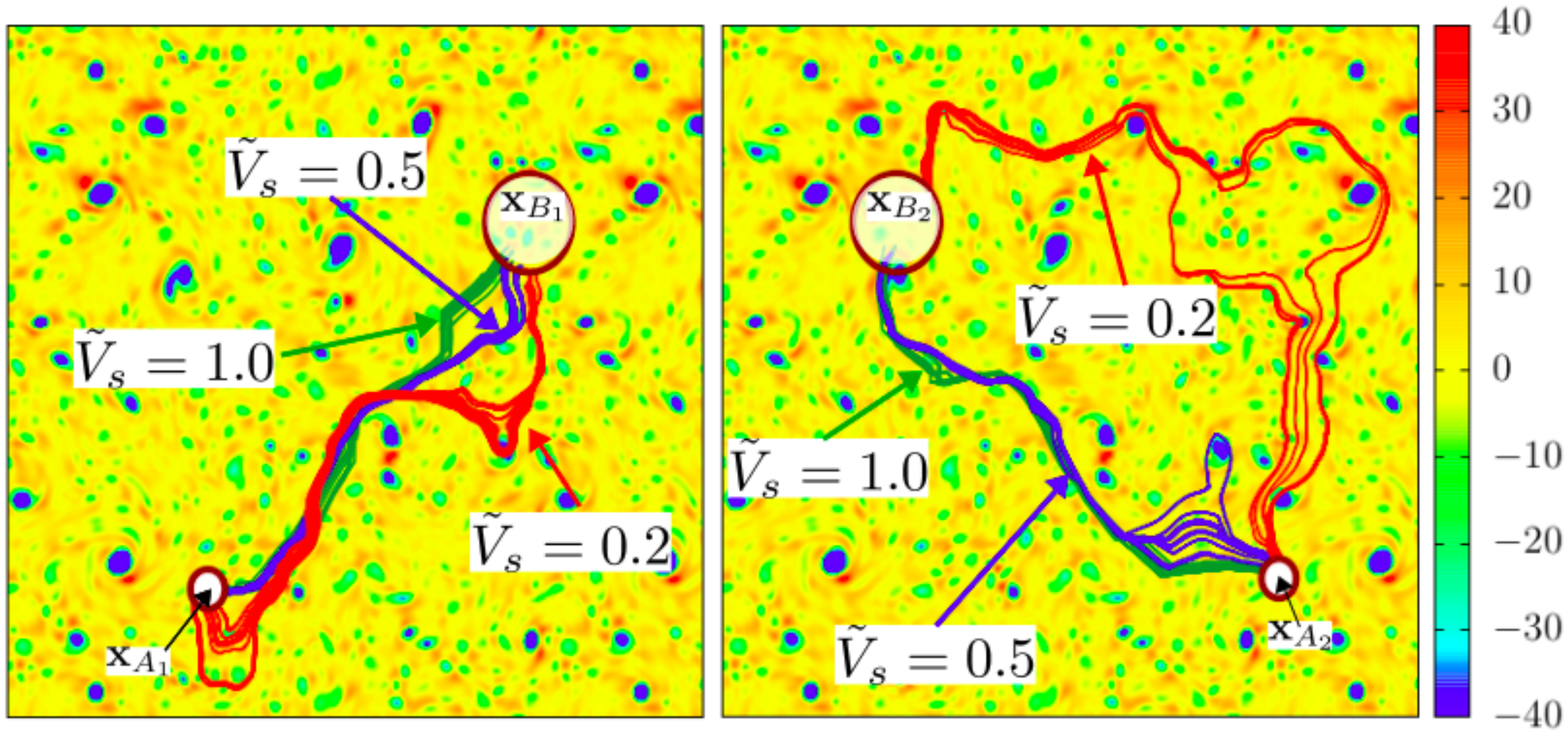}%
    \caption{Superposition of the trajectories of Fig.~\ref{fig:comp} with  {the Okubo-Weiss parameter $\Delta_{OW}$ (color coded in simulation units) as defined in Eq.~(\ref{eq:OW})} for P1 (left) and P2 (right).
    }%
    \label{fig:OW}%
\end{figure}
\begin{figure}%
    \includegraphics[scale=0.41]{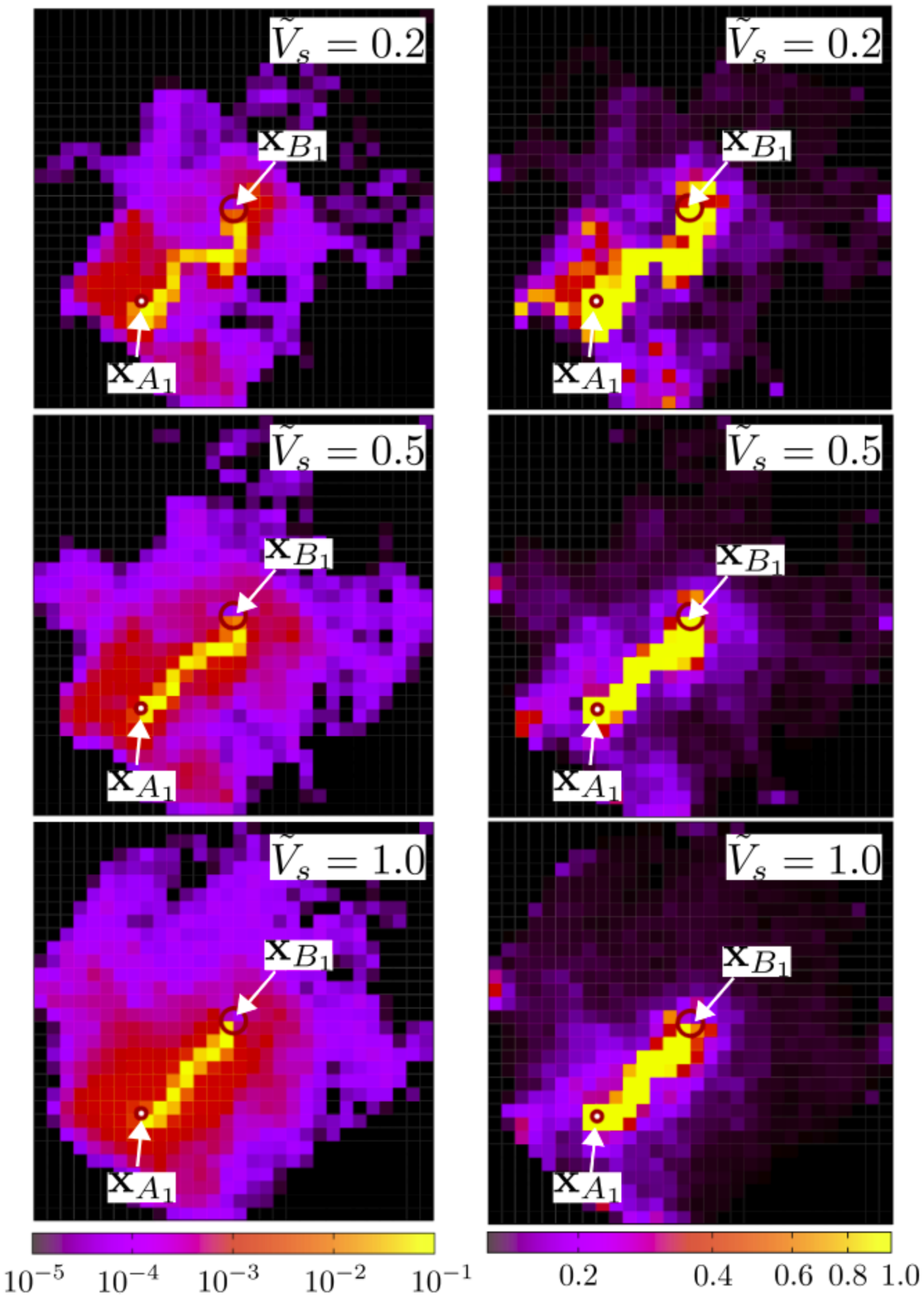}%
    \caption{(Left column) Color map of the density of visited states
    along the training phase for problem P1 and three different $\tilde V_{\rm s}$.
    (Right column) Degree of greediness of the best action for the {\it
    softmax} final optimal policy obtained by the RL for three different
    $\tilde V_{\rm s}$ and for the P1 point-to-point case. Similar results are
    obtained for P2 (not shown). }%
    \label{fig:matrixes}%
\end{figure}
\begin{figure*}%
\includegraphics[scale=0.6]{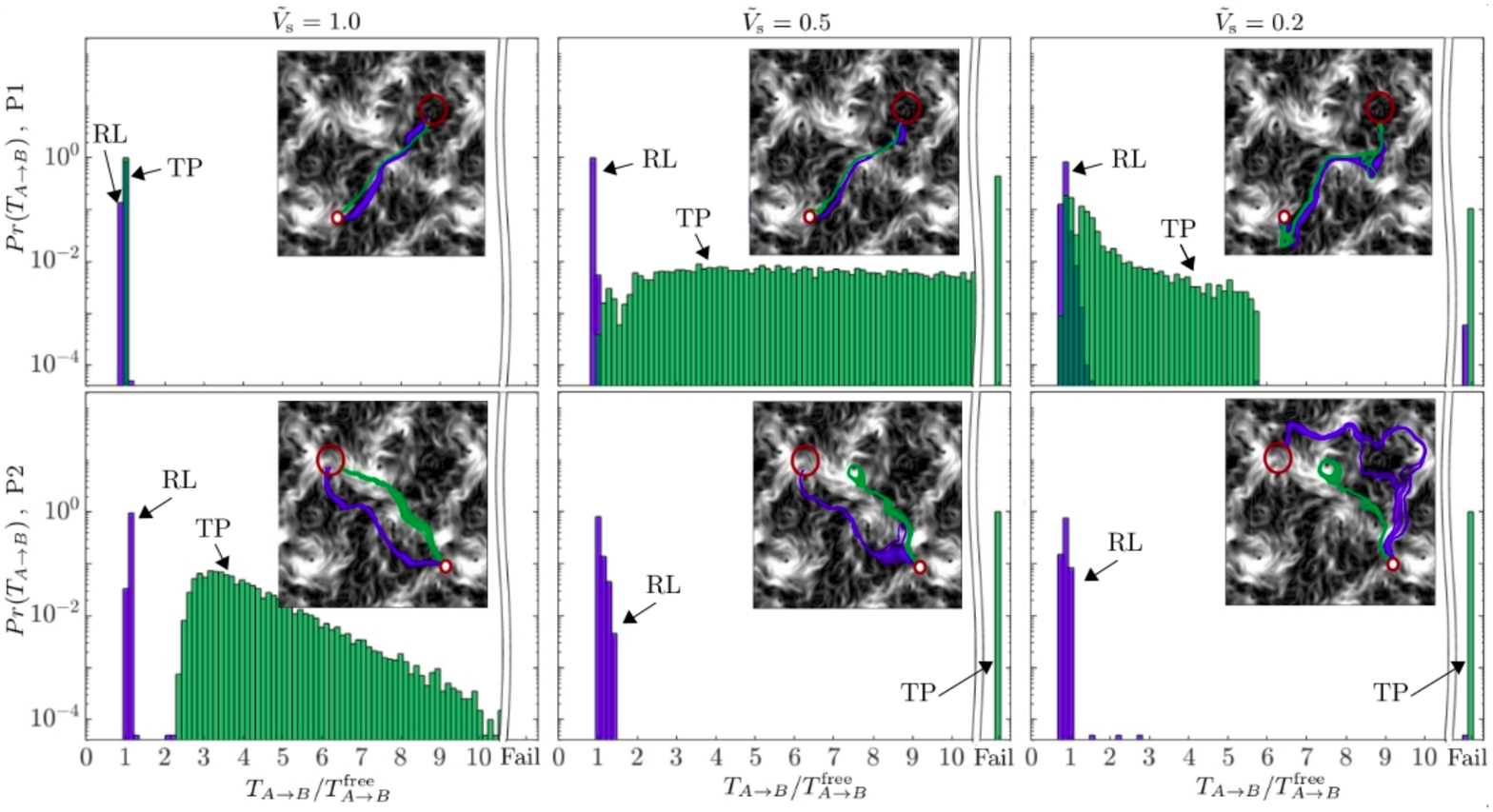}%
    \caption{Comparison between  {the arrival time histograms}, {Pr($T_{A \to B}$),} using the
    Reinforcement Learning (RL) (blue bars) and the Trivial Policy
    (TP) (green bars) with different values of $\tilde V_{\rm s}$ and for P1 (top row)
    and P2 (bottom row). 
    The bar on the right end of each panel denotes the probability of failure,  i.e. that the target is not reached.
    \label{fig:comparison} }%
\end{figure*}
\begin{figure*}%
    \includegraphics[scale=0.4]{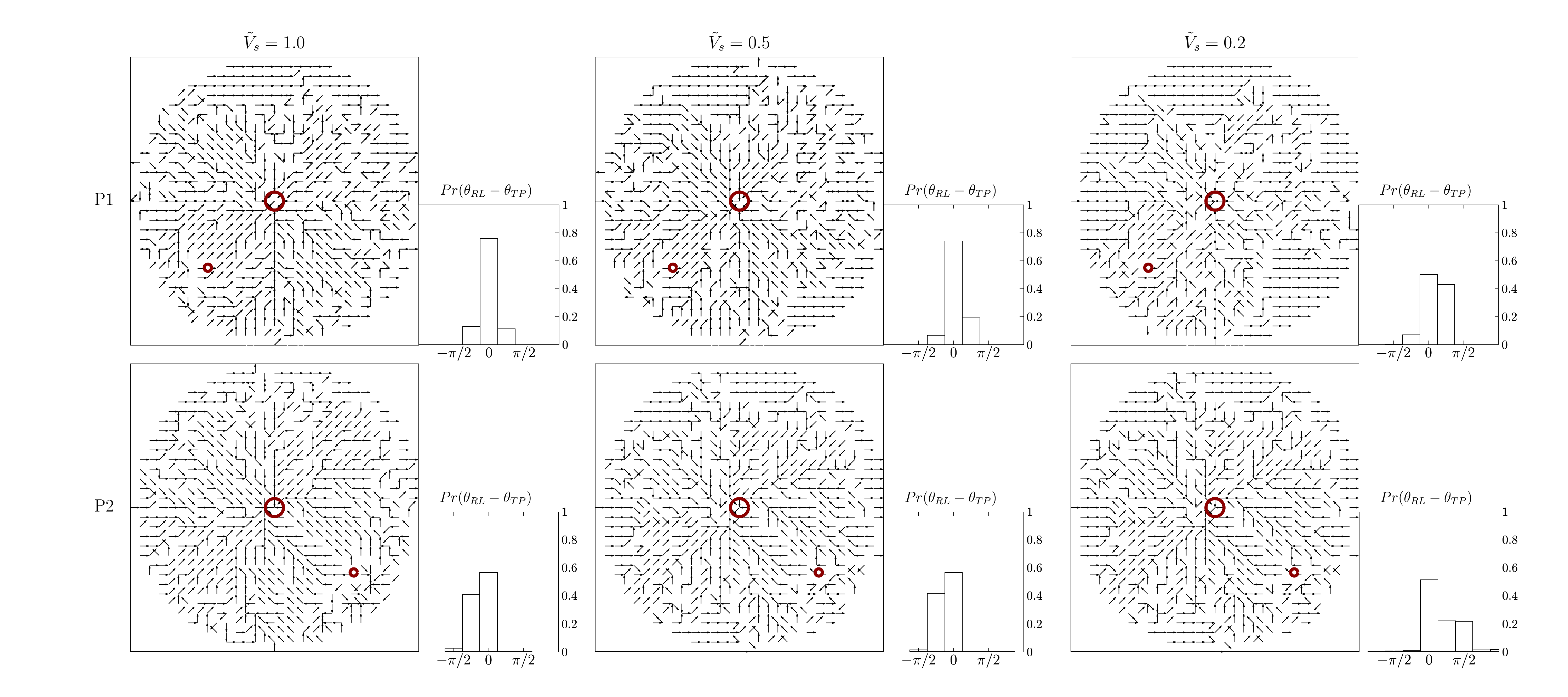}%
    \caption{2D map of the best actions selected by the RL for $\tilde V_{\rm s}=1,0.5,0.2$ (from left to right) and  for P1 (top row) and  P2 (bottom row). Next to each
    panel we show  the {histograms} of the angle mismatch, $Pr(\theta_{\rm RL} -\theta_{\rm TP})$,  between the greedy
    action selected by the RL and the action selected
    following the TP.
    }%
    \label{fig:map}%
\end{figure*}

\subsection{Results \LB{(time-independent flows)}}
\label{sec:RL_results}
\LB{Let us first examine the case when the flow field is time-independent.}
In Fig.~\ref{fig:training} we show the evolution of the total reward as a
function of the episode number for problem P1 using $\tilde{V}_{\rm s}=0.5$.
As one can see, the system reaches a stationary state and stable maximum
reward, indicating that the RL protocol has converged to a certain
policy. In the inset of Fig.~\ref{fig:training}, we show the trajectories of the
vessel following the policies extracted from three different
stages during the learning process. This illustrates that the policy evolves slowly toward one that generates stable and short paths joining $A_1$ and $B_1$. This is the first result supporting    the RL approach.

In Fig.~\ref{fig:comp} we show examples of trajectories generated with
the final policies for both problems P1 and P2 and for different values of $V_{\rm s}$.
For large slip velocities, $\tilde V_{\rm s}=0.5$ or
$\tilde V_{\rm s}=1.0$, the optimal paths are very close (but not identical) to
a straight line connecting the start and end points. For the case
with small slip velocity, $\tilde V_{\rm s}=0.2$, the vessel must
make use of the underlying flow to reach the target. This is particularly clear for problem P2 where it navigates on very {\it intense} flow regions, as can be seen by looking at the correlations
between the trajectories in red and the underlying flow intensity in the
right panel of Fig.~\ref{fig:comp}.  To further illustrate this, we superpose in Fig.~\ref{fig:OW} the example trajectories of Fig.~\ref{fig:comp} with the Okubo-Weiss \cite{okubo1970horizontal,weiss1991dynamics} parameter of the flow (defined as the discriminant of the eigenvalues of the fluid-gradient matrix $\ma A$)
\begin{equation}
   {\Delta_{OW} = } \left ( A_{11} - A_{22} \right )^2 + \left ( A_{21} + A_{12} \right
    )^2 - \left ( A_{21} - A_{12}  \right )^2\,.
\label{eq:OW}
\end{equation}
The sign of this parameter determines if the flow is straining (positive) or rotating (negative) and the magnitude determines the degree of strain or rotation.
When the slip velocity is small, $\tilde V_{\rm s}=0.2$ (red curves),  the vessel tends to get attracted to the vicinity of the rotating regions where {it exploit the  coherent head wind to reach the target quickly. }

One of the main results of this paper is connected to the high
robustness of the RL solution, especially if compared with the ON (see
later in Sect.~\ref{sec:comparison}). Here we want to show that this property is connected to the fact
that the RL optimal policy is the result of a systematic sampling of all
regions inside the flow, with information not restricted to the few
states that are visited by the shortest trajectories.  In the left
column of Fig.~\ref{fig:matrixes}, we show a color coded map for the
density of visited states for target P1 (similar results are obtained
for P2). As one can see, while there is obviously a high density close
to the optimal trajectory, the system has also explored large regions around it, allowing it to also store non-trivial information about neighbouring regions of the optimal trajectory.
Similarly, in the
right column of Fig.~\ref{fig:matrixes} we plot the degree of
greediness, defined as the probability of the optimal action for each
state:
\begin{equation}
g(s_i) = \max_{a_j} \pi^*(a_j,s_i)
\label{eq:greedy}
\end{equation}
in order to have a direct assessment of the randomness in the policy. Close to the optimal trajectory the policy is almost fully
deterministic, becoming more and more random as the distance to the optimal trajectory increases.

Finally,  we compared the optimal policies found
with RL to a {\it trivial} policy (TP), {where the angle selected at each $\Delta t$ is given by the action that points most directly towards the target among the $8$ different possibilities.}  In Fig.~(\ref{fig:comparison}) we show the trajectories
optimized with RL and the ones following the TP at different $\tilde
V_{\rm s}$ for problems P1 and P2, together with the {probabilities} of arrival times, $T_{A \to B}$. The TP is able to perform well only
when the navigating slip velocity is large, see  $\tilde V_{\rm s}=1$ for P1. Conversely, for the more interesting case when $\tilde V_{\rm s}$ is small, the TP produces many failed attempts (as illustrated by the bars to the right end of the histograms), the arrival times tend to be much longer { or infinite because the agents get trapped in recirculating regions from where it is difficult or impossible to exit.} The results of the TP are even more bleak in P2, where TP is only successful when $\tilde V_{\rm s}=1$ and for the other cases the vessels always get trapped in the flow.  In order to quantify the local differences between RL and TP along the optimal trajectories, we show in Fig.~\ref{fig:map} the greedy solutions {(solutions using the action with the highest probability)} selected  by the RL in the whole domain for all studied cases together with the probability of the angle mismatch between the greedy RL and TP, $Pr(\theta_{\rm RL} -\theta_{\rm TP})$,  (shown in the small boxes). {As one can see, there is always a certain  mismatch between the RL and the TP policies, confirming the difficulty to guess apriori the quasi-optimal solutions discovered by the RL.}
\LB{
\subsection{Results \LB{(time-dependent flows)}}
In this section we present some results for the more realistic, and more difficult, problem when we relax the assumption that the flow is slow or frozen and we consider Zermelo's problem for a time-dependent two-dimensional turbulent velocity field. The time evolution is obtained by solving the incompressible Navier-Stokes equations:
\begin{equation}
\label{eq:NS}
    \partial_t u_i + u_j \partial_j u_i = -\partial_i p + \nu \Delta u_i +f_i
\end{equation} 
on a periodic square with size $L=2\pi$ with $N=512^2$ collocation points, using a pseudo-spectral fully de-aliased code. In (\ref{eq:NS}), $p$ is the pressure, $\nu=1 \cdot 10^{-3}$ the viscosity and the forcing mechanism is Gaussian and delta-correlated in time, with support in Fourier space in the window $|{\bf k}| \in [5:6]$. A stationary statistics is achieved by adding an energy sink at large scale to stop the inverse energy cascade. The averaged spectrum is 
shown in the inset (a) of Fig. \ref{fig:time}.\\
In Fig.~\ref{fig:time} we show the learning curve for Zermelo's problem connecting the two locations $\bx_{A_3} \to \bx_{B_3}$ (see Fig. \ref{fig:timeevolution} for a graphical summary of the set-up) with propelling velocity $\tilde V_s=0.2$. In the inset (b) of Fig.~\ref{fig:time} we show the  evolution of the kinetic energy, $0.5 |{\bf u}(\bx,t))|^2$,   in the initial and final points for a time duration comparable to that of a typical navigation.
\begin{figure}%
    \includegraphics[scale=0.62]{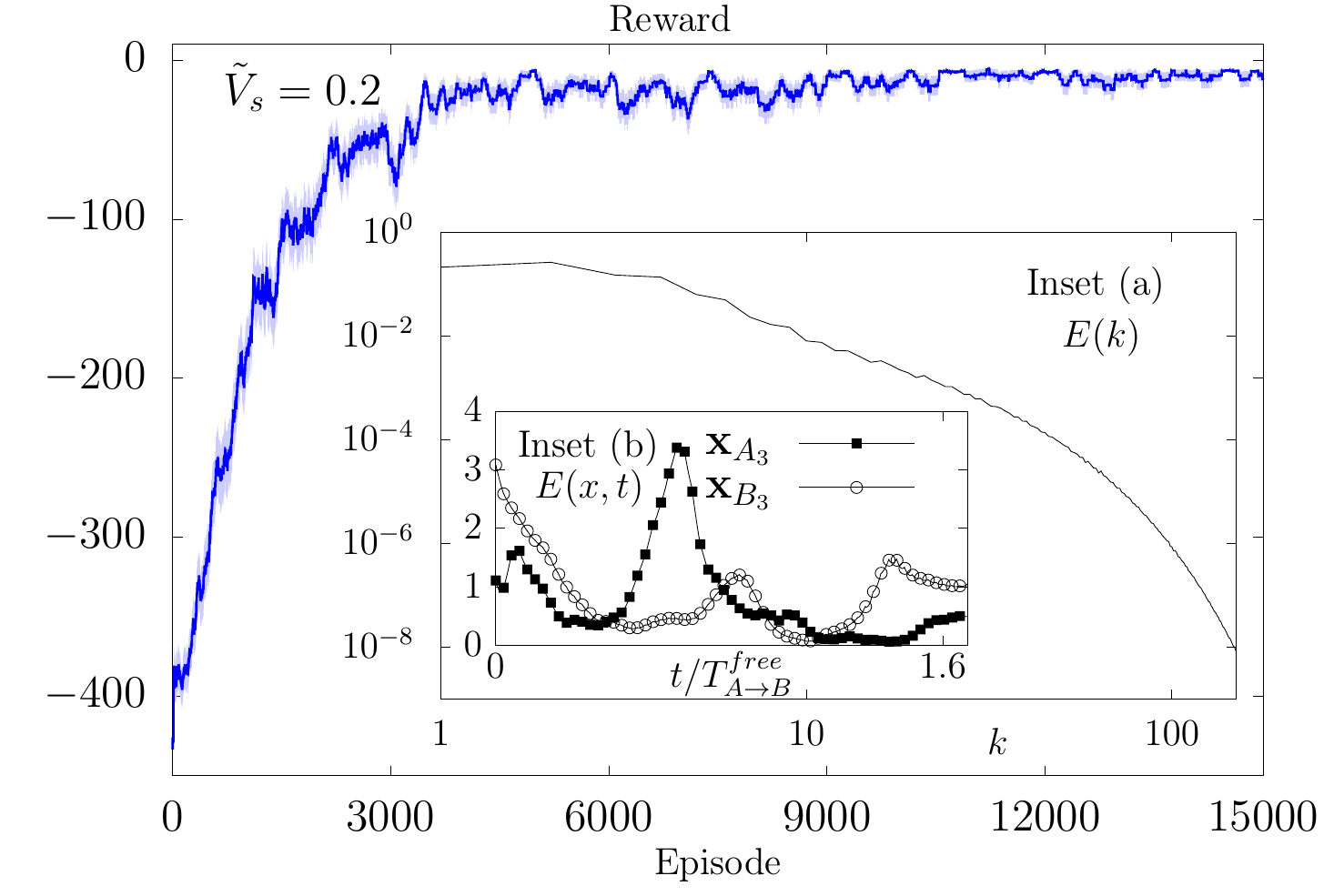}%
    \caption{\LB{Evolution of the total reward $r_{\rm tot}$ averaged over a window of $150$ consecutive  episodes versus the number of training episodes for $\tilde V_{\rm s} = 0.2$. The shaded area around the main curve indicates the standard deviation inside each averaged window. Inset (a): time-averaged energy spectrum for the 2d turbulent flow. Inset (b): kinetic energy evolution of the flow at the starting, $\bx_{A_3}$ (full squares) and at the final, $\bx_{B_3}$ (open circle).}
    }%
    \label{fig:time}%
\end{figure}
In Fig.~\ref{fig:timepdf} we show the PDF of arrival times, similar to Fig.~\ref{fig:comparison}, but for the time-dependent turbulent flow. We find that the RL approach clearly outperforms the trivial policy (TP), just as in the time-independent case. 
\begin{figure}%
    \includegraphics[scale=1]{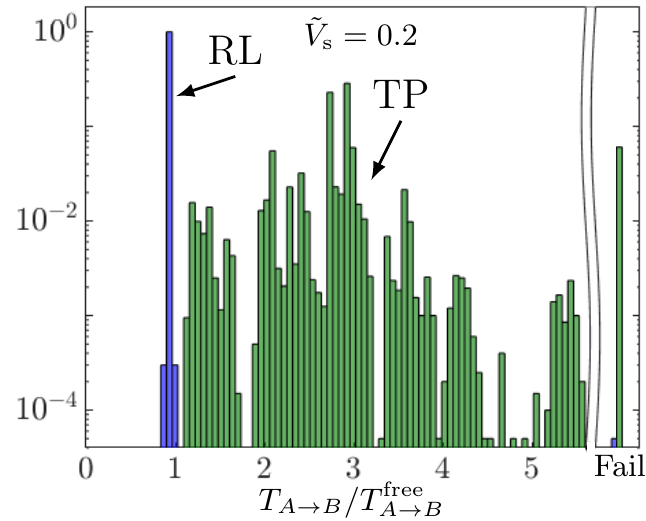}%
    \caption{\LB{Comparison between the arrival time histograms, {Pr($T_{A \to B}$),} using the
    Reinforcement Learning (RL) (blue bars) and the Trivial Policy
    (TP) (green bars) for $\tilde V_{\rm s}=0.2$ in the 2d turbulent time-dependent flow.
    }    
    }%
    \label{fig:timepdf}%
\end{figure}
The RL and TP policies are illustrated using six snapshots at different times in Fig.~\ref{fig:timeevolution}. The 6 snapshots show the time evolution of the flow and the growth of two sets of trajectories following learned or trivial policies.  
A video showing the full evolution can be found in the supplementary material.
\begin{figure*}%
    \includegraphics[scale=0.6]{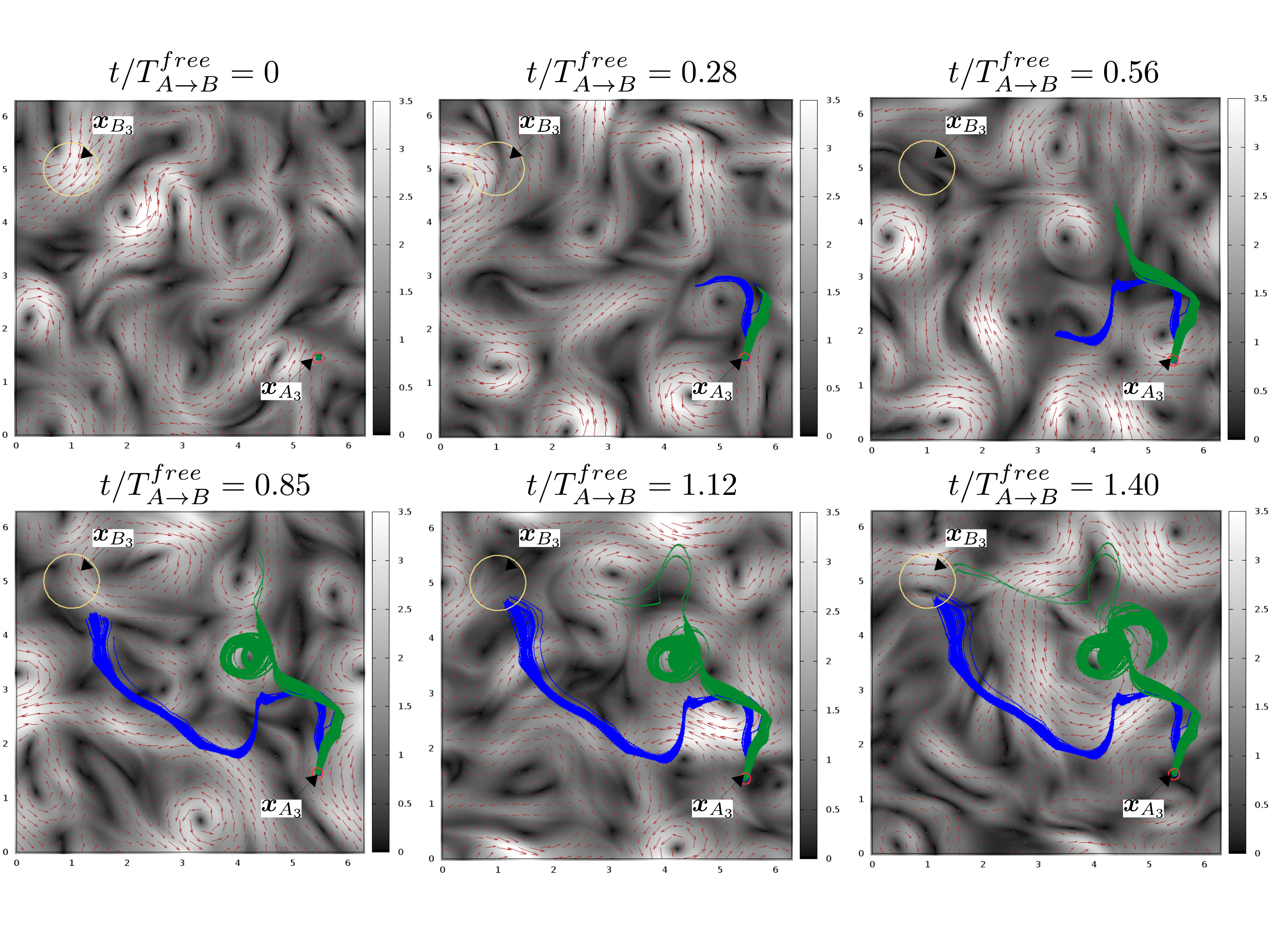}%
    \vspace{-15pt}
    \caption{\LB{Time-dependent case: the six panels contains six snapshots in time taken during evolution of the flow and of two sets of ten trajectories one following the trivial policy (TP) (green lines) and the other following the RL policy (blue lines). These figures have been extracted from the movie available in the supplementary material.}
    }
    \label{fig:timeevolution}%
\end{figure*}
Furthermore, in Fig. \ref{fig:kinetic} we show the evolution of the flow kinetic energy along a representative vessel trained with RL compared with the equivalent case following a TP trajectory. As one can see, the RL policy is able to take advantage of the {\it good} flow structures to accelerate toward the target and to avoid those that would bring it away, at difference from what happens to the TP which blindly falls inside trapping or accelerating vortical structures. 
\begin{figure}%
    \includegraphics[scale=0.6]{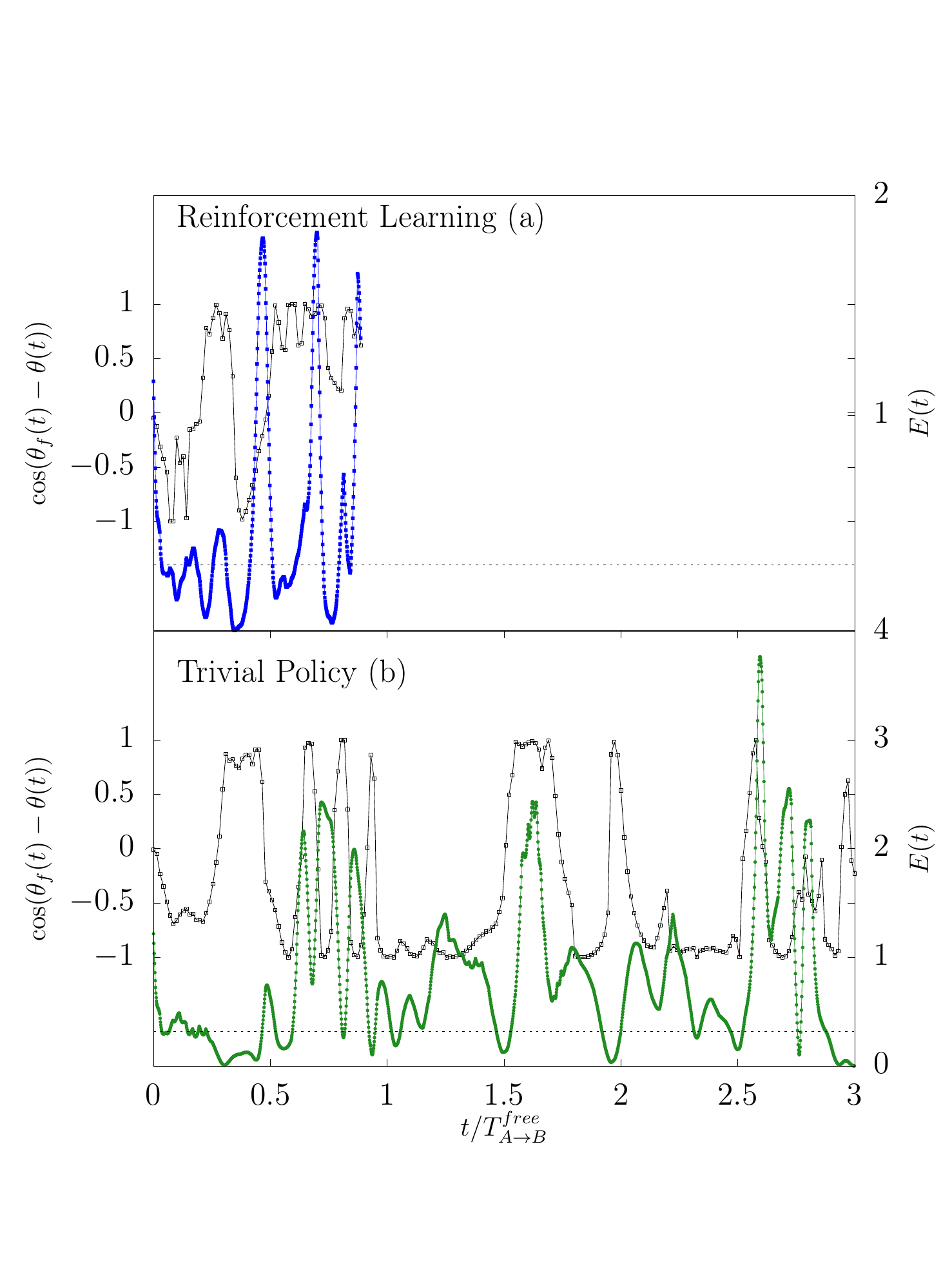}%
    \vspace{-15pt}
    \caption{\LB{Two representative evolution of the kinetic energy measured along the trajectories of two vessels, one following the RL policy (blue colour panel a) and one following a TP (green colour panel b). The horizontal dashed lines represent the fixed kinetic energy of the autonomous vessel propulsion, $0.5 V_s^2$. In the same figures the cosine of the difference between the navigation direction $\theta(t)$ and the flow direction $\theta_f(t)$ is shown as reported on the left y-axes. Notice the ability of the RL policy to take advantage of regions with strong flow energy to reach the target earlier.}
    }
    \label{fig:kinetic}%
\end{figure}
}
\LB{In conclusion, there are no doubts that RL is able to converge on quasi-optimal navigation policies even in flows with spatio-temporal chaos as the case of multi-scale 2d turbulent flows as here analyzed. Moreover, discovered policies are far from the trivial option to navigate-by-eyes and steering always in the target's direction. }
\section{The Optimal Navigation approach}
\label{sec:ON}
\LB{We explore optimal navigation only for the time-independent flow, because then the problem can be solved analytically~\cite{bryson2018}.
In this section we implement the approach proposed in Ref. \cite{bryson2018} for the time-independent case of Fig.~\ref{fig:initial}.
}

\subsection{Methods}
\label{sec:ON_methods}
Eq.~(\ref{eq:ON}) gives the evolution of the steering angle
that minimizes the time it takes to navigate from $\ve x_A$ to $\ve x_B$ provided that the system starts at the optimal initial angle, as was first derived by Zermelo \cite{zermelo1931}. Following \cite{bryson2018}, this
control strategy can be obtained by mapping the problem onto a classical
mechanics problem with a Hamiltonian
\begin{align*}
    {{\cal H}(\ve X_t,\ve p_t,\theta_t)=[\ve u(\ve X_t)+V_{\rm s}\ve n(\theta_t)]\cdot\ve p_t-1\,,}
\end{align*}
where {$\ve p_t=(p^x_t,p^y_t)$} are the generalized momenta of {$\ve X_t$}.  The
corresponding Hamiltonian dynamics become
\begin{align}
    \label{eq:HX}
    \dot{\ve X}_t&=\frac{\partial H}{\partial \ve p}=\ve u(\ve X_t)+V_{\rm s}\ve n(\theta_t)\\
    \label{eq:Hp}
    \dot{\ve p}_t&=-\frac{\partial H}{\partial\ve X}=-\ma A^{\rm T}(\ve X_t)\ve p_t\\
    0&=\frac{\partial H}{\partial\theta}=V_{\rm s}[-p^x_t\sin\theta_t+p^y_t\cos\theta_t]\,,
    \label{eq:Htheta}
\end{align}
where $\ma A$ is the fluid gradient matrix with components
$A_{ij}=\partial_ju_i$ {evaluated along $\ve X_t$}.  By construction, using the principle of least
action, solutions from $\ve x_A$ at $t=0$ to $\ve x_B$ at $t=T_{A\to B}$ of this
dynamics are extreme points of the action evaluated along a trajectory
(\ref{eq:theta}):
\begin{align*}
    \int_{0}^{T_{A\to B}}\hspace{-0.3cm}{\rm d}t{\cal L}=\int_{0}^{T_{A\to B}}\hspace{-0.3cm}{\rm
    d}t\left[\dot{\ve X}_t\cdot\ve p_t-{\cal H}\right]=\int_{0}^{T_{A\to B}}\hspace{-0.3cm}{\rm
    d}t=T_{A\to B}\,.
\end{align*}
Thus, the trajectory with the optimal time to navigate to the target
satisfies the dynamics (\ref{eq:HX}--\ref{eq:Htheta}).
Eq.~(\ref{eq:Htheta}) gives $\ve p\propto\ve n$ and it follows that the
time-optimal control is obtained by solving the joint
equations~(\ref{eq:HX}) and ~(\ref{eq:Hp}) using $\ve n=\ve p/|\ve p|$.
The corresponding angular dynamics, {$\dot\theta_t={\rm d}\, ({\rm atan}(p^y_t/p^x_t))/{\rm d}t$}, is identical to Eq.~(\ref{eq:ON}).
We remark that the dynamics~(\ref{eq:Hp})  tend
to orient the vessels transverse to the maximal stretching directions
of the underlying flow.  As a consequence, vessels avoid
high strain regions and accumulate in vortex regions.


Solutions of Zermelo's problem deliver  the quickest
trajectory joining the starting point to the target point for each
initial condition. The challenge lies in finding the initial
direction that hits the target point in the shortest time. In the set-up
described in Section~\ref{sec:setup}, the target is not a single point
but instead an area. We view this as a continuous family of Zermelo's
problems, where each target point corresponds to a point in the target
area with an optimal initial angle and corresponding optimal time.
Thus, the optimal path corresponds to the solution of Zermelo's problem that has the quickest trajectory.

As we shall see, it is not straightforward to find the optimal trajectory by refining the angle {$\theta_{t_0}$} for the initial condition. The complication  is that Zermelo's dynamics are often unstable in a non-linear flow.  This implies that even initial conditions very close to each other may end up at different locations in the flow.
As a consequence it is hard to find the optimal strategy: local refinement of $\theta_{t_0}$ tend to result in local minima rather than the global one.

\subsection{Results}
\label{sec:ON_results}
\begin{figure*}%
    \includegraphics[width=18cm]{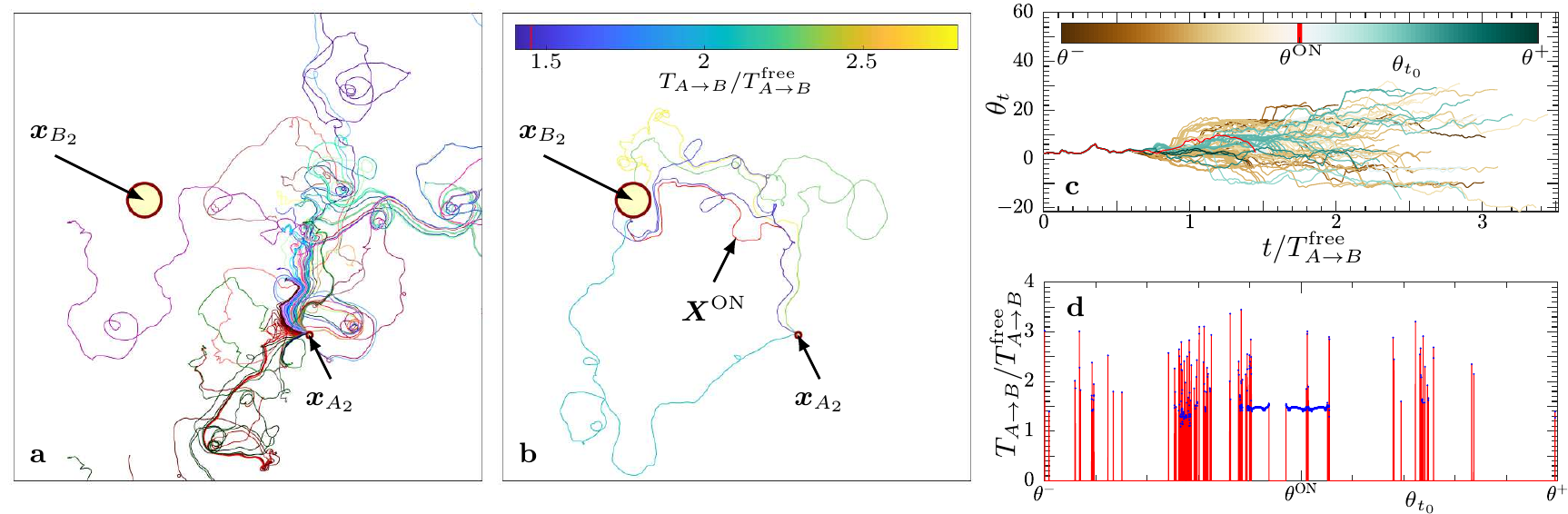}%
    \caption{(a): visualization of {$100$} trajectories with
    starting angle uniformly selected on a grid between $0$ and $2\pi$
    and constant velocity $\tilde V_{\rm s} = 0.2$. Trajectories are distinguished using {different} colors. (b): The subsample of
    trajectories that reaches the target (out of $10000$) are shown. The
    color map indicates the time $T_{A\to B}$ when the target is reached.
    The quickest trajectory, ${\bf X}^{ON}$,  out of 10000 is colored red and defines the optimal initial angle, $\theta^{\rm ON}\approx 2.4193$ {for $\tilde V_{\rm s}=0.2$ and P2}.
    (c): Evolution of $\theta_t$ for target-reaching trajectories starting close to $\theta^{\rm ON}$. Trajectories are color coded according to their initial angle $\theta_{t_0}$ and terminated when the target is reached. Out of 10000 trajectories with uniformly equispaced initial angles in $\theta^-\le\theta_{t_0}\le\theta^+$ with $\theta^\pm=\theta^{\rm ON}\pm \pi/5000$, $14\%$ reach the target and are displayed.
    (d): Time $T_{A\to B}$ to reach the target as a function of $\theta_{t_0}$ for the trajectories in panel (c). Zero time corresponds to trajectories not reaching the target. The time is plotted discretely (blue points) and continuously (red line) to indicate transitions between successes and failures.
 }%
 \label{fig:ON_traj}
\end{figure*}
\begin{figure}%
    \includegraphics[width=8cm]{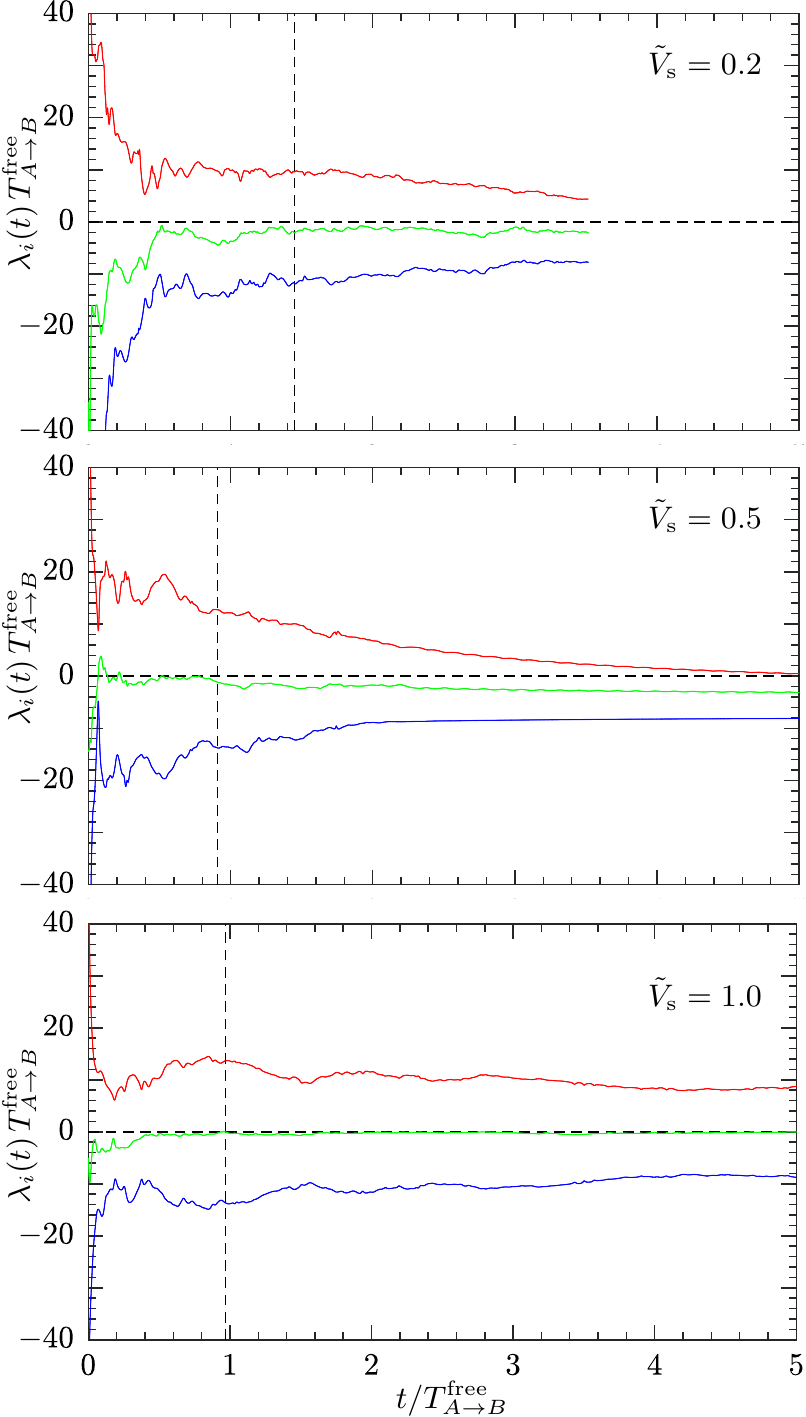}%
    \caption{FTLE for the quickest ON trajectories for target P2 (starting at $\ve x_{A_1}$ with initial angle $\theta^{\rm ON}$) for three different navigation speeds, $\tilde V_{\rm s}$. The vertical dashed line denotes the time the target is reached.
 }%
 \label{fig:Lyap}
\end{figure}
To study this problem, we integrate Eqs.~(\ref{eq:ON}) and (\ref{eq:theta}) numerically using a fourth-order Runge-Kutta scheme with a small time step, $\delta t=10^{-4}$ for 100 time units.
We have explicitly checked that reducing the time step to $10^{-5}$ does not change vessel trajectories significantly on the time scales considered in this work.
We consider the same targets and values of $V_{\rm s}$ as for the RL.
In Fig.~\ref{fig:ON_traj}a we show the results for ON running {100} trajectories with uniformly gridded initial
values of $\theta$ between $0$ and $2\pi$ and $\tilde V_{\rm s} = 0.2$ (for target P2). As one can see, {the trajectories wander around the flow, all missing the target.} Fig.~\ref{fig:ON_traj}b shows a repetition of the experiment for a larger number of trajectories (10000) and only the {few trajectories} that reach the target are shown (and terminated at the target). In order to empirically understand the stability of the protocol, we identified the optimal initial angle $\theta^{\rm ON}$ as the trajectory in Fig.~\ref{fig:ON_traj}b that reaches the target in the shortest time (red) and run another 10000 trajectories with initial angles in an interval of length $\pi/2500$ around $\theta^{\rm ON}$ (this is the interval of so far unexplored initial conditions around $\theta^{\rm ON}$).
Fig.~\ref{fig:ON_traj}c shows the evolution of $\theta_t$ for those trajectories that reach the target.
We observe that there is a wide variability in the time to reach the target and that the trajectories are well mixed in the long run: the order of $\theta_t$ does not reflect the initial ordering $\theta_{t_0}$ at the time scale when the target is reached.
Fig.~\ref{fig:ON_traj}d summarizes which initial angles reach the target and which fail.
We observe that successes depend intermittently on the initial angle: continuous bands of successful initial conditions are interdispersed with regions of failures.
We also observe that within this sample there exist trajectories that reach the target quicker than the trajectory starting at $\theta^{\rm ON}$.
However, due to the intermittent nature of successes, it is not possible to continuously change $\theta_{t_0}$, starting at $\theta^{\rm ON}$, to find the best sampled trajectory without passing regions of failure. This highlights the problem in refining the initial condition to find the global optimal trajectory in the ON protocol to the non-linear system considered here.

As shown in Fig.~\ref{fig:ON_traj}, ON is able to produce trajectories
that join the starting and ending points. We will compare their times
with the ones coming from the RL methods in the next section. Here we instead make a more quantitative analysis of the stability of the ON solutions. In order to do this, we evaluate the FTLE along
a phase-space trajectory $(x,y,\theta)$ governed by the system of
Eqs.~\eqref{eq:theta} and~\eqref{eq:ON}. The FTLE are the local
stretching rates, $\lambda_i(t)$ for $i=1,2,3$, of a small phase-space
separation $\ve w \equiv (x-x',y-y',\theta-\theta')$.  By polar decomposition, we can write $\ve
w(t)=\ma V(t)\ma R(t)\ve w(0)$, where $\ma R(t)$ and $\ma V(t)$ are
rotation and positive definite stretching matrices.
The FTLE $\lambda_i(t)$ are defined from the eigenvalues $\exp[t\lambda_1(t)]$,
$\exp[t\lambda_2(t)]$ and $\exp[t\lambda_3(t)]$ of $\ma V$.  If the
maximal FTLE is positive, the ON trajectory is unstable and small
deviations from the trajectory are exponentially amplified with time.
In Fig.~\ref{fig:Lyap}, we show results of the FTLE for the quickest trajectories reaching the
target using ON, i.e. trajectories starting at $\theta^{\rm ON}$ as defined in Fig.~\ref{fig:ON_traj}b. We find that the maximal FTLE is positive for the time it takes to reach the target for all velocities
$\tilde V_{\rm s}$ considered. On the other hand, for times large enough, many trajectories approach fixed-point attractors, leading to a smooth decay towards
negative FTLE. This effect becomes more evident the smaller the navigating velocity $V_{\rm s}$ is.
For $V_{\rm s} \rightarrow 0$ the system of equations~\eqref{eq:theta} decouples from Eq.~\eqref{eq:ON} and the
dynamics cease to be sensitive to small perturbations.

In conclusion, Fig.~\ref{fig:Lyap} shows that the system is unstable to small perturbations on the time scales needed to reach the target. This explains the observed behaviour of the trajectories in Fig.~\ref{fig:ON_traj} and why the ON approach is untractable for practical applications.

\section{Comparison between RL and ON}
\label{sec:comparison}

In this section we make a side-by-side comparison of the  RL and ON approaches.
To better highlight the RL stability compared to the ON solution we need to specify how the two
sets of simulations are initialized. While the RL initial conditions are
chosen in a circle of radius  $d_{A_i}$ centered at $\bx_{A_i}$ with a
small spread in the initial angles (typically the probability of the non-greedy actions  which is in the initial state  of the order of $0.01$), we initialize the ON simulations at the exact spatial
starting point, $\bx_{A_i}$, and with uniformly distributed initial angles in an interval of length $0.02$ around $\theta^{\rm ON}$ as defined in Fig.~\ref{fig:ON_traj}b. The ON approach could not be initialized starting in a box of side length $d_{A_i}=0.2$ because its unstable dynamics prevents it to work if the range of initial conditions is too wide.

\begin{figure}%
    \includegraphics[width=9cm]{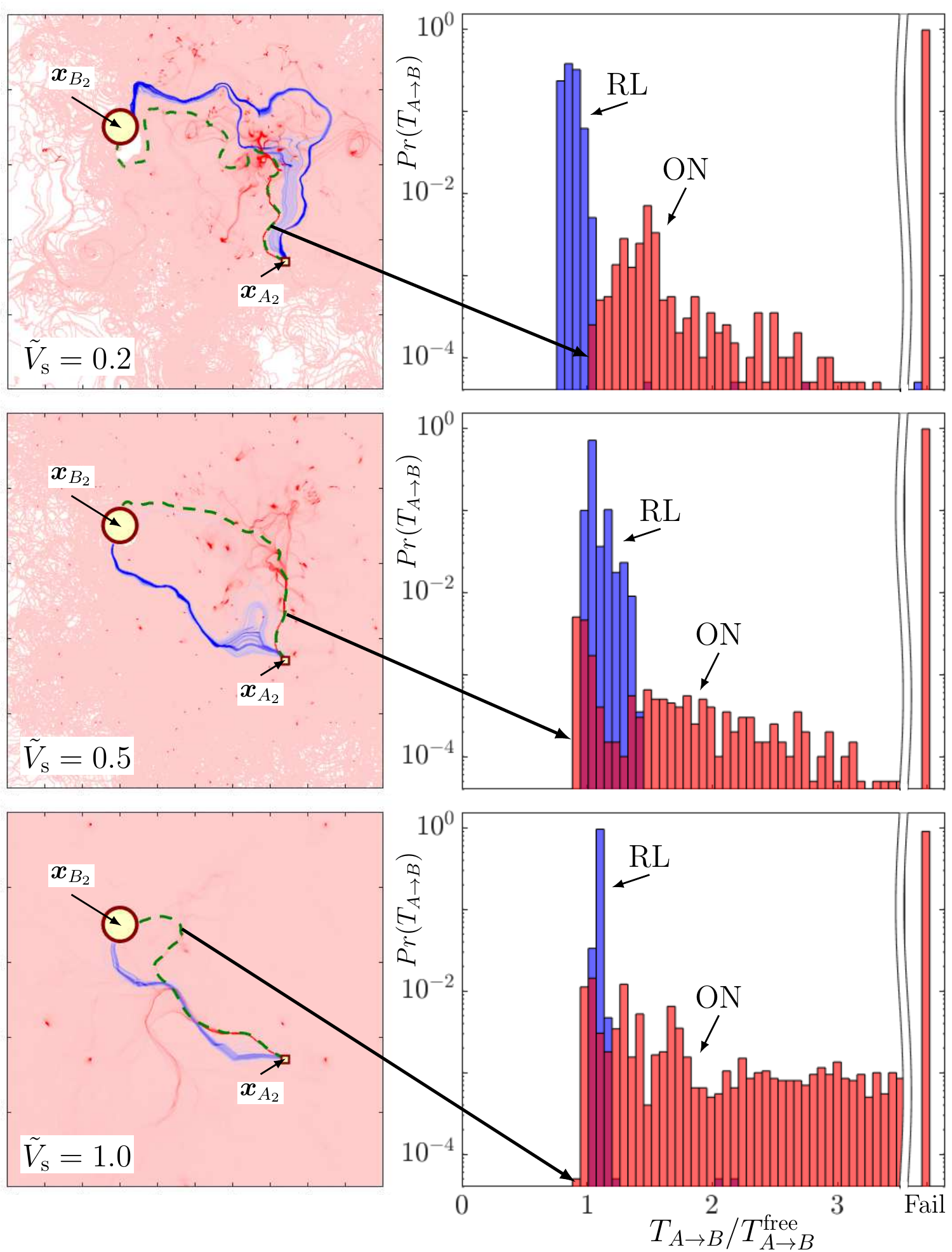}%
    \caption{Left column: spatial concentrations of trajectories for three values of $\tilde
    V_{\rm s}$. The flow region is  color coded proportionally to the
    time the trajectories spend in each pixel area for both  ON  (red) and  RL (blue). Light colors refer to low occupation and bright to high
    occupation.  The green-dashed line shows the
    best ON out the $20000$ trajectories. Right column: { arrival time histograms} for ON (red) and RL (blue). {Probability of } not reaching
    the target within the upper time limit is plotted in the {\it Fail} bar.
    }
    \label{fig:ONvsRL}
\end{figure}

We find that the minimum time taken by the best trajectory to reach the target is of the same order for the two methods. The main difference between RL and ON lies instead
in their robustness.
We illustrate this by plotting the spatial density of trajectories in the left column of
Fig.~\ref{fig:ONvsRL} for the optimal policies of ON and RL with three values of $\tilde V_{\rm s}$ and initialized as described above.
We observe that the RL trajectories (blue colour area) follow a coherent region in space, while the ON trajectories (red colour
area) fill space almost uniformly, especially for large values of
$\tilde V_{\rm s}$. Moreover, for small navigation velocities, many trajectories in the ON system approach regular attractors, as visible by the high-concentration regions. Similar results are found for the
trajectories following problem P1 (not shown).

The right column of Fig.~\ref{fig:ONvsRL} shows a comparison between the {probability} of arrival times for the trajectories illustrated in the
left column. This provides a quantitative estimation
of the better robustness of RL compared to ON: even though the
ON best time is comparable or, sometimes, even slightly smaller than the RL minimum time, the ON
{probability} has a much wider tail towards large arrival times and it is
always characterized by a much larger number of failures. All these
results highlights a strong instability of the ON approach.\\

\section{Conclusions}
\label{sec:conclusions}

We have presented a first systematic investigation of Zermelo's time-optimal navigation problem in a realistic
2D turbulent flow. We have developed a RL approach based on an actor-critic algorithm, which is able to find quasi-optimal discretized policies with strong robustness against changing of the initial condition.  In particular, we have considered constant navigation speeds, $V_{\rm s}$, not exceeding the maximum flow velocity $u_{\max}$, down to values of $V_{\rm s} \sim 0.2 \, u_{\max}$ and for all cases we successfully identified quick navigation paths and close to optimal policies that are strongly different from the trivial choice to navigate towards the target at all times.  We have also implemented a few attempts with an  additive noise  in the equations for the vessel evolution and found that RL is able to reach a solution even for this case (results will be reported elsewhere).  Furthermore,  we investigated the relation between the optimal paths and the underlying topological properties of the flow, identifying the role played by coherent structures to guide the vessel towards the target. Finally, \LB{for the time-independent flow} we have compared RL with the Optimal Navigation approach showing that the latter exhibits a strong sensitivity on the
initial conditions and is thus inadequate for real-world applications.
{Many potential applications can be envisaged. \\
\LB{It is important to stress that the RL approach implemented here requires information about the flow evolution for the duration of the optimal trajectory. In realistic problems of travel planning, this information must be provided by a model of the evolution of the flow. Furthermore, each flow evolution and each couple of starting/arrival points requires individual optimization, it is highly unlikely that a solution to one optimization problem can be robustly applied to other situations. In this work we implemented RL using an actor-critic structure because of its flexibility to be extended to Deep-RL approaches \cite{mnih2015human,zhu2017target,arulkumaran2017deep,novati2018deep}.
Other RL algorithms, such as Q-learning \cite{sutton2018}, could also be implemented. Our approach is based on repeated trials for an important reason: the objective of Zermelo’s problem is to find the trajectory that connects two points in the minimal time, not just any trajectory. For this reason we must be able to run different iterations and find the minimum (or at least a local minimum). Nonetheless, if we change our objective to just connecting two points, our RL approach can be extended to a fully online one where only one agent swims around the flow continuously until it reaches the target. We plan to tackle that problem in future research.} For future work, it is also key to probe the efficiency of the different approaches considered here for 3D geometries, where already the simple uncontrolled tracer dynamics, $V_s=0$, can be chaotic even in time-independent flows \cite{dombre1986chaotic,bohr2005dynamical}, opening new challenges for the optimization problem. Moreover, similar optimal navigation problems can be reformulated for inertial particles \cite{toschi2009lagrangian,gibert2012small,bec2007heavy,mordant2001measurement}, where the control is moved to the acceleration with important potential applications to buoyant geophysical probes \cite{roemmich2009argo}. Work in these directions is in progress and will be reported elsewhere. }\\
\section{Supplementary material}
Please see supplementary material for a movie showing the side-by-side evolution of two bunches of trajectories evolved following the optimal RL or the trivial policy in a time-dependent 2d turbulent flows:
\url{https://www.fisica.uniroma2.it/~biferale/MOVIES/ZermeloMovie.arXiv1907.08591.mp4}

\section{Acknowledgments}
We acknowledge A. Celani for useful comments. L.B., M.B. and P.C.d.L. acknowledge funding from the European Union Programme (FP7/2007-2013) grant  No.339032. K.G. acknowledges funding from the Knut and Alice Wallenberg Foundation, Dnar. KAW 2014.0048.

\bibliography{bibliography}

\end{document}